\documentclass[aps,pre,footinbib,showpacs,twocolumn]{revtex4-1}
\usepackage{amssymb,latexsym,mathrsfs}
\usepackage{amsfonts}
\usepackage{amsmath}
\usepackage{graphicx}
\usepackage{ulem}

\newcommand{\beq}{\begin{equation}}
\newcommand{\eeq}{\end{equation}}
\newcommand{\beqn}{\begin{eqnarray}}
\newcommand{\eeqn}{\end{eqnarray}}
\newcommand{\bearr}{\begin{array}}
\newcommand{\enarr}{\end{array}}

\newcommand{\ra}{\rangle}
\newcommand{\la}{\langle}

\newcommand{\id}{\textrm{d}}

\def\bea{\begin{eqnarray}}
\def\eea{\end{eqnarray}}
\def\ba{\begin{array}}
\def\ea{\end{array}}
\def\n{\nonumber}

\def\ve{\varepsilon}

\usepackage{color}

\begin{document}

\title{Negative Differential Mobility in Interacting Particle Systems}
\author{Amit Kumar Chatterjee$^1$} \email{amit.chatterjee@saha.ac.in} 
\author{Urna Basu$^2$ and P. K. Mohanty$^1$}
\affiliation{$^1$CMP Division,  Saha Institute of Nuclear Physics, HBNI, 1/AF Bidhan Nagar, Kolkata 700064, INDIA}
\affiliation{$^2$LPTMS, Université Paris-Sud 11, Orsay, France}

\begin{abstract}
Driven  particles in  presence of  crowded environment, obstacles or kinetic constraints  often exhibit negative differential mobility (NDM) due to their decreased  dynamical activity. We propose a new mechanism for complex many-particle systems where slowing down of certain {\it non-driven} degrees of freedom by the external field can give rise to NDM. This phenomenon, resulting from  inter-particle interactions, is illustrated in a pedagogical example of two interacting random walkers, one of which is biased by an external field while the same field only slows down the 
other keeping it unbiased. We also introduce and solve exactly the steady state of several driven diffusive systems, including a two species exclusion  model, asymmetric misanthrope and zero-range processes, to show explicitly that this  mechanism  indeed  leads to NDM. 
\end{abstract}

\maketitle

The linear response of a system close to thermal equilibrium is characterized by the fluctuation-dissipation theorem \cite{Kubo}. The current generated by a small external drive can be predicted from equilibrium correlation functions using the so-called Green-Kubo relations \cite{Green,Kubo2} and the mobility, that is the ratio of the average particle current to the external force, remains necessarily positive. Away from equilibrium, the linear response formula gets modified \cite{Baiesi2009} and positivity of mobility is no longer guaranteed. In fact, particles driven far away from equilibrium might show absolute negative mobility \cite{anm1,anm2,anm3,anm4} or negative differential mobility.

Negative differential mobility (NDM) refers to the physical phenomenon when current in a driven system 
decreases as the external drive is increased \cite{zia,rev}. This has been observed in various systems, both in context of particle \cite{zia,Barma,Jack} and thermal transport \cite{Li,Hu,deb}. In particular, the occurrence of NDM  of driven tracer particles  in presence of obstacles \cite{Dhar,neg,neg2,Baiesi2015,Franosch} or in steady laminar flow \cite{Sarracino2016} or crowded medium \cite{Saraccino,Benichou2016} have been studied extensively in recent years.
NDM has also been observed in driven many-particle systems in presence of kinetic constraints \cite{Sellitto} or obstacles \cite{Baiesi2015, Reichhardt}. The emergence of NDM in all these systems is typically associated with `trapping' of the driven particles; the  obstacles or the crowded environment slows   down the particle motion as the driving is increased, which, in turn, reduces the current. This trapping is usually characterized by a decrease in the so called `traffic' or dynamical activity \cite{neg,neg2}, which plays a key role in understanding response of nonequilibrium systems \cite{Baiesi2009,Baiesi13}.

 In  interacting systems, current  constitutes of contributions from many degrees of freedom (or different types of particles or modes), each of which can be driven separately; the  external field might also act differently on different modes. Reducing the dynamical activity or `traffic' of  some degrees  of freedom which are not necessarily driven, can be expected to influence the  dynamical activity of the driven ones due to interaction; this raises a possibility of having a new mechanism to induce NDM in interacting   systems.

In this article we  propose  a generic mechanism for NDM in driven interacting particle systems.
We  show  that if  the inter-particle  interaction slows down the  time scale of  modes which are not necessarily driven, then a non-monotonic behaviour of current might arise leading to a negative differential mobility. 
We introduce an interacting two-species  model to demonstrate explicitly that the increased external bias  on  one species can in fact  reduce the  dynamical activity of the  other. NDM appears to be a  direct consequence of this effect; trapping of driven  particles alone by the increased  bias may  not be  sufficient. To validate this scenario we introduce and study several exactly solvable models, the simplest  being two distinguishable random walkers on a  one dimensional lattice  interacting via mutual exclusion. We explicitly show  that   when one of  the  particles is driven by an external field, corresponding current may  decrease  when the escape rate of the second, undriven, particle decreases as a function of the field. We generalize this scenario to  interacting many particle systems with two  or more species of particles, with and without hardcore interactions.

Let us first consider a system of interacting particles of two species on a one-dimensional periodic lattice of 
size $L.$ Each site $i=1,2\dots L$  is either vacant or  occupied   by  at most  one  particle  of  kind $A$ 
or $B$,  represented  by the site variables  $\tau_i=0,A,B$ respectively.  The configuration of the system evolves following the dynamical rules,
\beq
A0 \mathop{\rightleftharpoons}_{q}^{p} 0A ~;~X B0 \mathop{\rightarrow}^{r(X)} X0B  ~;~0BX \mathop{\rightarrow}^{l(X)} B0X, \label{eq:XOB1}
\eeq
where $r(X), l(X)$  are the hop rates of a $B$ particle to the right and left empty neighboring sites, respectively, when $X=0,A,B$ is the occupancy of the other neighbor.  Here we consider a specific case $q=1-p,$  $r(0)=w$ and all other  rates  $r(X) = 1-w = l(X);$ this corresponds to the physical scenario where isolated B particles are driven by a constant bias quantified by $w$ while other $B$ particles diffuse symmetrically.  
In addition to the dynamics \eqref{eq:XOB1}, we also allow an exchange between neighboring $A$ and $B$ particles,
\beq
AB \mathop{\rightleftharpoons}_{\alpha}^{\alpha} BA\label{eq:XOB2}
\eeq 
to  ensure ergodicity in the phase space. Clearly the  dynamics conserves both densities $\rho_{A,B}= N_{A,B}/L$  where  $N_{A,B}$ are the number of particles of the respective species. 
For a  fixed value of parameters $w,\alpha$  and densities $\rho_{A,B}$  we  would like to see  how  the current  
varies with  $p$ or  equivalently the biasing field $\ve= \ln \frac{p}{1-p},$ defined consistently with local detailed balance \cite{kls}, taking  $k_BT=1$ for the sake of simplicity. Note that $B$ particles are not driven by this external field $\ve.$ 
\begin{figure}[t]
 \centering
 \includegraphics[width=8.7 cm]{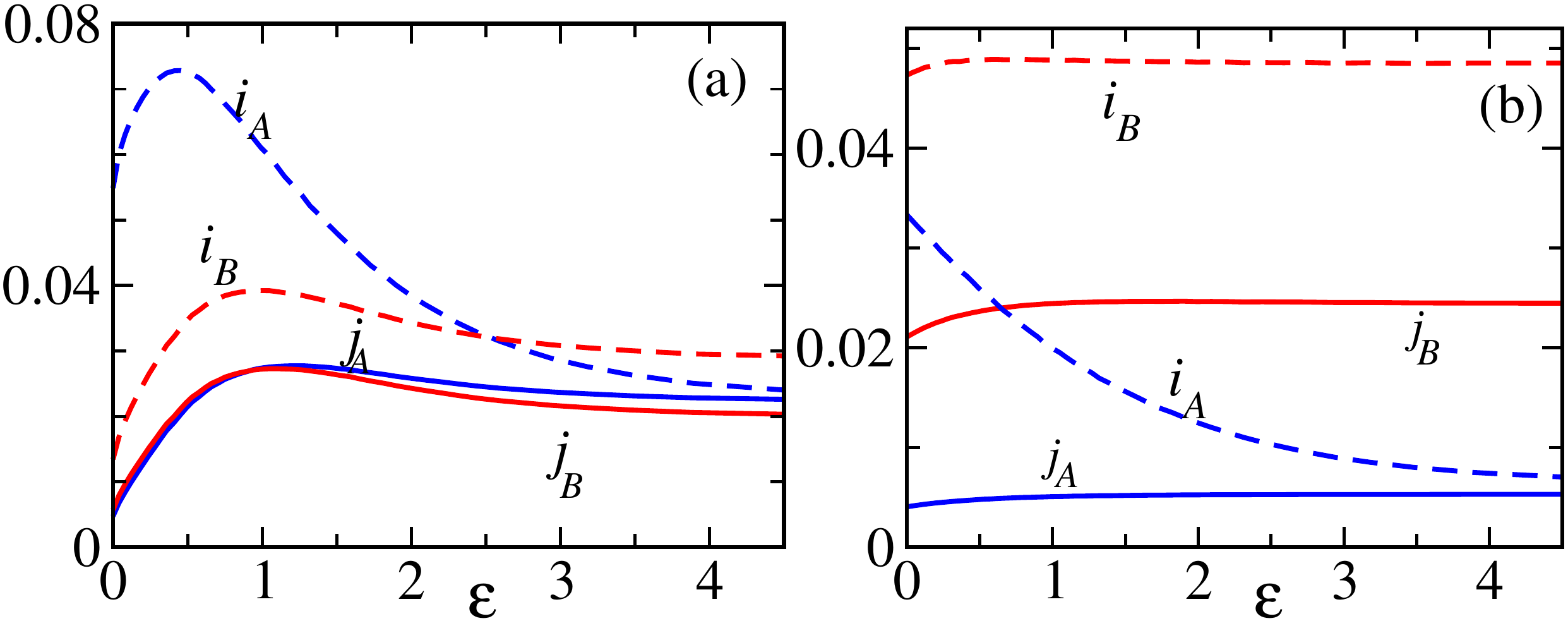}
 \caption{Current $j_{A,B}$ (solid line)  and `traffic' $i_{A,B}$ (dashed line), from simulations of the  two species model defined in Eqs.~(\ref{eq:XOB1}) and \eqref{eq:XOB2} for $L=1000, \rho_A=0.1,$ $w=0.9$ and $\alpha=0.01.$ (a) and (b) correspond to  $\rho_B=0.1,0.5$ respectively.}\label{fig:XBO}
\end{figure}
Using  Monte Carlo simulations  of the model  we  measure  the average current $j_{A,B}= (\la n_r^{A,B}\ra - \la n_l^{A,B}\ra)/L$  where $\la n_{r}^{A,B}\ra$ ( $\la n_{l}^{A,B}\ra$)  are  the average number of right (left)  jumps  of $A$ or $B$ particles in  unit time, measured in the steady state. 
Figure  \ref{fig:XBO} shows   $j_{A,B}$ for two different densities $\rho_B=0.1,0.5$ of the $B$-particles; NDM is seen in the former case while the current shows a monotonic behaviour in the latter one. To understand the origin of the NDM in this system and to investigate if the driving is, somehow, leading to a `trapping' of the particles, we measure the average `traffic' or time-symmetric current  $i_{A,B} =(\la n_r^{A,B}\ra + \la n_l^{A,B}\ra)/L$ of the $A,B$ particles which are also shown in the Fig.~\ref{fig:XBO} (dashed lines). The inverse of the traffic measures the typical time-scales associated with particle jumps; it seems that $i_A$ is a decreasing function of the drive signifying that the $A$ particles, apart from being driven, are indeed also `slowed down' by the field $\ve.$  Surprisingly, however, it turns out that, decreasing $i_A$ alone is not sufficient (see Fig. \ref{fig:XBO}(b)) rather it is the slowing down of the $B$ particles which is the decisive factor giving rise to NDM \cite{note1}.
 To observe NDM, decreasing traffic of the driven degrees is  certainly necessary  (as observed in  other many-particle systems  \cite{Sellitto,Baiesi2015}) but its insufficiency here  
indicates presence of possible additional controlling  factors. 

Based on this phenomenological picture we propose a possible mechanism for NDM in interacting systems: particle current in a driven many particle system might show a non-monotonic behaviour if some  modes, which are not driven by the external field, slow down with  increased driving. To  substantiate this scenario, we consider several interacting particle systems 
where some  degrees of  freedom  are  biased  by  the external  field, whereas  the same field  slows down other  modes explicitly. The aim  is  to  show from exact steady state calculations that 
particle current in such situations indeed exhibit NDM, purely due to the effect of inter-particle interaction. 

{\it Two random walkers:} 
As a simple prototypical example  of  two  interacting  current-carrying modes  we consider two distinguishable particles, denoted by $A$ and $B,$  on a periodic  lattice interacting via hardcore exclusion, i.e.,   the occupancy  of  the site $i$ is  $\tau_i=A,B,0$ where  particles $A$ and $B$ cannot occupy the same site. 
 The particles follow  a dynamics,
\bea
A0 \mathop{\rightleftharpoons}_{q}^{p} 0A, \quad B0 \mathop{\rightleftharpoons}_{\psi}^{\psi} 0B.\label{eq:dy_two_rw}
\eea     
In this Two Random Walkers (TRW)  model, the external bias  affects  the  particles differently: the $A$ particle is driven by the  external field  $\ve= \ln(p/q)$  whereas the $B$ particle performs an unbiased random walk  with jump rate $\psi(\ve)$ --- also depending on the field $\ve.$

\begin{figure}[t]
 \centering
 \includegraphics[width=8.8 cm]{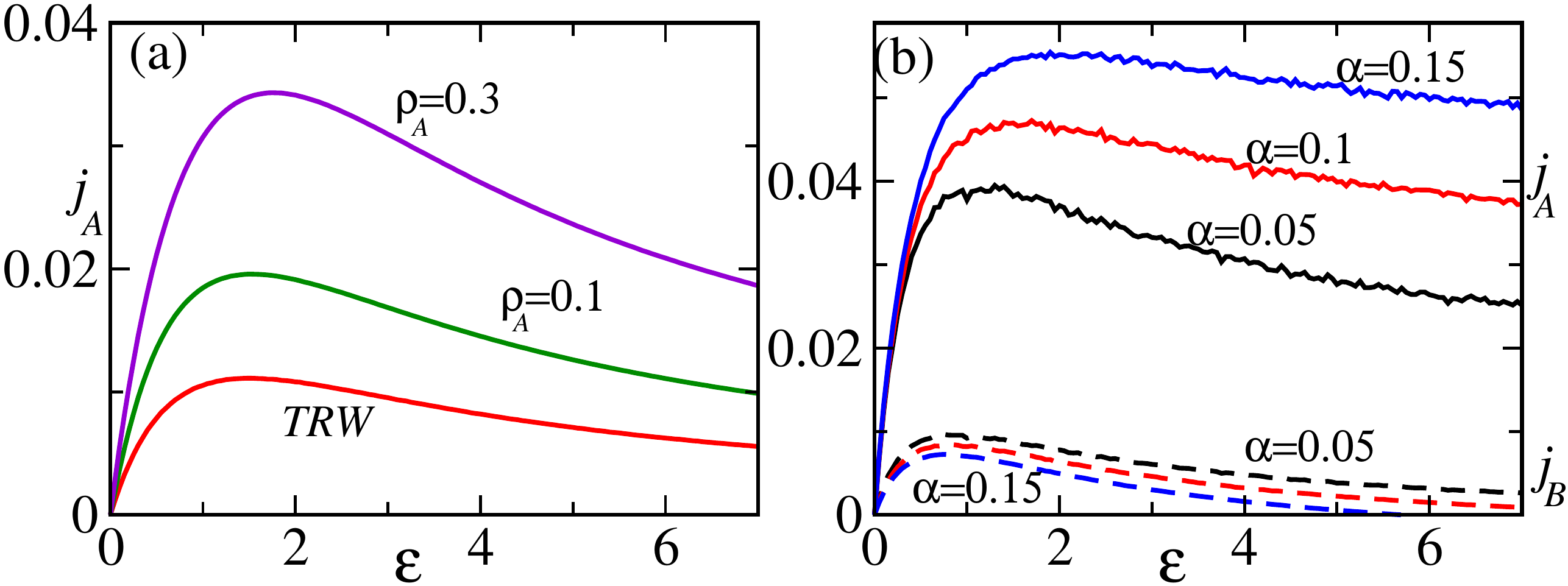}
 \caption{NDR in interacting two-species model. (a) $j_{A}$ obtained from exact steady state results  for  
 $\alpha=0$  are  shown for  $\rho_B=\rho_A=0.1, 0.3;$ the threshold values are $\ve^*=1.53, 1.76$ respectively. 
 For TRW model we plot $j_A/200$ for better visibility. 
 (b) For $\alpha>0,$  $j_A$  (solid lines) and $j_B$ (dashed lines) are  obtained from simulations with  
 $L=1000, \rho_A=0.1, \rho_B=0.3,$ and  different $\alpha=0.05,0.1$ and $0.15.$}
\label{fig:AB}
\end{figure}

We use the so called Matrix Product Ansatz \cite{MPAevans} to express the steady state weight  $P(C)$  for  any configuration $C$ in a matrix product form : $P(C)= Tr[\prod_{i=1}^{L}X_i]$ where the matrix $X_i=\hat A\delta_{\tau_i,A}+ \hat B\delta_{\tau_i,B}+\hat E\delta_{\tau_i,0}$  represents occupancy $\tau_i$ of  the site $i.$  The matrices $\hat A,\hat B,\hat E$ must satisfy the following set of algebraic relations to satisfy the Master equation in the steady state
\bea
&&{\hat A}^{2}=0={\hat B}^{2};~ p{\hat A} \hat E- q \hat E {\hat A}=x_0 {\hat A}; \cr && \psi( {\hat B} \hat E- \hat E {\hat B})=x_0 {\hat B},
\label{eq:mat_al_two_rw}
\eea
where $x_0$ is an auxiliary scalar. We find a $2\times2$ representation of the matrices  ${\hat A},{\hat B},\hat E$  which  satisfies the algebra \eqref{eq:mat_al_two_rw} with a choice $x_0=\psi\frac{p-q}{\psi+q},$  
\bea
\begin{array}{ccc}
 {\hat A}=\left( \begin{array}{cc} 0 & 1 \\ 0 & 0 \\ \end{array} \right) &  {\hat B}=\left( \begin{array}{cc} 0 & 0 \\ 1 & 0 \\ \end{array} \right) & 
\hat E=\left( \begin{array}{cc} \gamma & 0 \\ 0 & 1 \\ \end{array} \right),
\end{array}
\label{eq:mat_rep_two_rw}
\eea
where $\gamma=\frac{p+\psi}{q+\psi}.$ The partition function of the system of size $L$  is then  $Z_L=\sum_C P(C)=  Tr[ ({\hat A}+{\hat B}+\hat E)^L]=\frac{\gamma^{L-1}-1}{\gamma-1}.$  The average stationary current of $A$ particle  is  
\beq
j_A=p \langle A0\rangle - q \langle  0A \rangle = x_0\frac{Z_{L-1}}{Z_L}\simeq 
\frac{(p-q)\psi}{(p+\psi)},
\label{eq:cur_A_two_rw}
\eeq
we have taken the thermodynamic limit $L \to \infty$ in the last step. Due to the presence of the interaction, the $B$ particle also exhibits a stationary current which depends on $\ve;$ in fact, $j_B =j_A$ (as expected in the  absence of particle exchange) and  the total particle current $j =  j_A + j_B =2 \frac{(p-q)\psi}{(p+\psi)}.$
The response of the current $j$ to a small increase in $\ve$ is quantified by the differential mobility
\bea
\frac{\id j}{\id \ve}= \frac{(p'-q')\psi + (p-q)\psi'}{(p+\psi)} - \frac{(p-q)\psi(p'+\psi')}{(p+\psi)^2},
\eea
where prime denotes derivative w.r.t. $\ve.$ Equilibrium corresponds to $\ve=0,$ i.e., $p=q$ (remember that $q=e^{-\ve}p$) and the mobility, as expected near equilibrium, is positive irrespective of the functional form of $\psi.$ On the other hand, in the large driving limit $\ve \to \infty,$ assuming $p$ and $p'$ remain finite, we have,
\bea
j'(\infty) = -\lim_{\ve \to \infty} \frac{p^2\psi^2}{(p+\psi)^2} \frac{\id}{\id \ve}\left(\frac 1p + \frac 1\psi \right)
\eea
This sets a criterion for NDM in TRW model: if asymptotically the  increasing rate of $\psi^{-1}$ is  larger than the  decreasing rate   of $p^{-1},$ then there   will be a finite  bias $\ve^*>0$ above  which 
the  response is negative. Since the inverse rates measure the diffusion time scales, and the particles  here interact  via  strong repulsive interaction (here hardcore), this criterion  is in tune with the  proposition given  in  Ref. \cite{Saraccino}. In particular, for the cases $q=1/(1+e^{\ve})$ (i.e., $p=1-q$) or $q=e^{-\ve}$ ($p=1$), any choice of  $\psi(\ve)$ for which   $\psi'(\infty) <0$  would exhibit NDM.  Current $j_A=j/2$ as a function of $\ve$  for $q=e^{-\ve}, \psi(\ve)  = 1/(1+\ve)$  is shown in Fig. \ref{fig:AB}(a); NDM occurs here for $\ve>\ve^*= 1.505.$ 

A somewhat similar model with two coupled Brownian particles is studied  in Ref.~\cite{ac} where only one of the particles was driven by both a static and a time-periodic force and it was observed that the second particle  exhibits NDM in some parameter regime. In contrast,  here we show that NDM occurs for  both the particles when the escape rate of the undriven particle is decreased with increasing  drive.

The TRW model, being a  simple  prototype  of  two  current  carrying modes,  lacks 
some important features of  realistic   driven systems. In the following we study several more 
complex driven interacting systems and  show that a similar mechanism indeed induces NDM.

{\it Two species exclusion process:} 
Our next  example is   
a generalized version of dynamics (\ref{eq:dy_two_rw}) with an  added particle  exchange  dynamics 
\bea
A0 \mathop{\rightleftharpoons}_{q}^{p} 0A, \quad B0 \mathop{\rightleftharpoons}_{\psi}^{\psi} 0B, 
\quad AB \mathop{\rightleftharpoons}_{\alpha}^{\alpha} BA.
\label{eq:dy_many_rw}
\eea
Unlike the TRW model, now  we  have macroscopic numbers  of  $A$ and $B$ particles, with
conserved  densities $\rho_A$ and $\rho_B$ respectively.   

First let us consider the case $\alpha=0.$ In absence of    
particle exchange  the  number of $A$s ($B$s) trapped between to consecutive $B$s ($A$s) are conserved 
and thus the configuration space  is not ergodic. One can however choose to work  in one particular sector; the  dynamics  then enforces ergodicity within that sector. Let  us work in  a sector with exactly one A particle between two consecutive  $B$s; the configurations   are  now  $C\equiv\{A 0^{n_1}B 0^{n_2}A 0^{n_3}B 0^{n_4}\dots B 0^{n_{2M}}\},$ each  having  exactly $N$  number of $A$s and  $B$s, and  $\sum_{i}n_i= L-2N$ number of vacancies. 
The  steady state weights of  this particular equal density ($\rho_A= \frac NL=\rho_B$) sector, 
can be  obtained exactly  using matrix product ansatz, by representing  $A,B,0$ 
as matrices  ${\hat A},{\hat B},{\hat E}$ respectively. The required matrix algebra in this case 
turns  out  to be same as   Eq.~\eqref{eq:mat_al_two_rw}   indicating Eq.~\eqref{eq:mat_rep_two_rw} as 
a possible representation. However, unlike the TRW model, here  we need to deal with finite densities  $\rho_{A,B}.$ 
The  grand canonical partition function  is now $Z_L=  Tr[(z {\hat A} + {\hat B}+{\hat E})^L]$  
where $z$ is the fugacity associated with only A particles (additional   fugacity for B particles  is not needed  as $\rho_B=\rho_A$). In  the thermodynamic limit, $Z_L= \lambda_+(z)^L$ where $\lambda_\pm(z)=\frac12(1+\gamma \pm \sqrt{ (1-\gamma)^2+ 4 z})$ are the eigenvalue of $(z {\hat A} + {\hat B}+\hat E)$ and $\gamma = (p+\psi)/(q+\psi).$ This leads  to  $\rho_A= z\frac{\id}{\id z} \ln \lambda_+(z)=
z/(\lambda_+^2 -\lambda_+\lambda_-).$
The steady state current   of $A$ particles is now
\beq
j_A= z(p \langle  A0\rangle- q\langle 0A\rangle)= \frac{z (p-q \gamma)}{ \lambda_+^2(\lambda_+ - \lambda_-)}\label{eq:cur_many_rw}
\eeq
Explicit calculation shows  that the current of $B$ particles  $j_B =\psi (\langle B0\rangle-\langle 0B\rangle)$ is same as $j_A$ (as expected for $\alpha=0),$ thus $j= 2 j_A.$ 
The differential response $\frac{\id j}{\id \ve}$ becomes negative 
as the field $\ve$ is increased beyond some threshold $\ve^*,$ which depends  on  the densities $\rho_A=\rho_B$ 
as shown in Fig. \ref{fig:AB}(a) for  $p=1, q=e^{-\ve},$ and $\psi(\ve) =  1/(1 + \ve).$

For $\alpha>0,$ we do not have an exact solution; however, Monte Carlo simulation confirms that the model still exhibits NDM in a fairly large  range of particle densities. Fig. \ref{fig:AB}(b) shows plots of $j_A$ and $j_B$ versus $\ve$ for different values of $\alpha$ for  $\rho_A=0.1, \rho_B= 0.3.$

{\it Asymmetric Misanthrope Process:} 
It is interesting to ask whether it is possible to see NDM in systems without hardcore exclusion. To this end we investigate an asymmetric misanthrope process (AMP) \cite{AMP} on a one-dimensional lattice where each site $i$ can hold any number of particles $n_i \ge 0.$ The particles can hop  to their  right or left  nearest  neighbors  with a rate  that depends on the occupation of both departure  and arrival sites, 
\beq
\{ n_i, n_{i+1}\}  \mathop{\rightleftharpoons}^{u_r(n_i,n_{i+1})}_{u_l( n_{i+1}+1,n_i-1)} \{n_i-1, n_{i+1}+1\};
\label{eq:AMP_basic}
\eeq
the functional form of the  rate functions $u_{r,l}(.),$ for right and left hops  are different.
This dynamics conserves density $\rho = \sum_i n_i/ L.$ The asymmetric rate functions correspond to driving fields $E_{mn}= \ln \frac{u_r(m,n)}{u_l(n+1,m-1)}$ acting on bonds with local configurations $(m,n).$
 Clearly, if $E_{mn}=0 ~~ \forall~ m,n,$ we have $u_r(m,n) = u_l(n+1, m-1)$ and
the system is in equilibrium satisfying  detailed balance condition with all configurations being equally likely.

We now choose a set of specific rate  functions, 
{\small
\beq
u_r(m,n) = \left\{ \begin{array}{ll} \psi & n=0\\ 1  & n>0\end{array}\right.;~
u_l(m,n) = \left\{ \begin{array}{ll} \psi & m=1\\ e^{-\ve} & m>1,n=0\\ \frac12 & m>1,n>0
\end{array}\right.,\nonumber
\eeq
}
which corresponds to
\beq
 E_{mn}=\left[\ln 2 + (\varepsilon-\ln2)\delta_{m,1} \right] (1-\delta_{n,0}). \label{eq:AMP}
\eeq
Here, rightward hopping of  particles to  vacant  neighbors are not biased (as  both the rightward hop 
and corresponding reverse hop occur with same rate $\psi$) whereas other rightward jumps are biased by an 
external field which depends on the  occupation of the departure site: 
$\ve$ when the departure site has only one  particle or otherwise  a constant field $\ln 2$. 

\begin{figure}[t]
 \centering
 \includegraphics[width=8.8 cm]{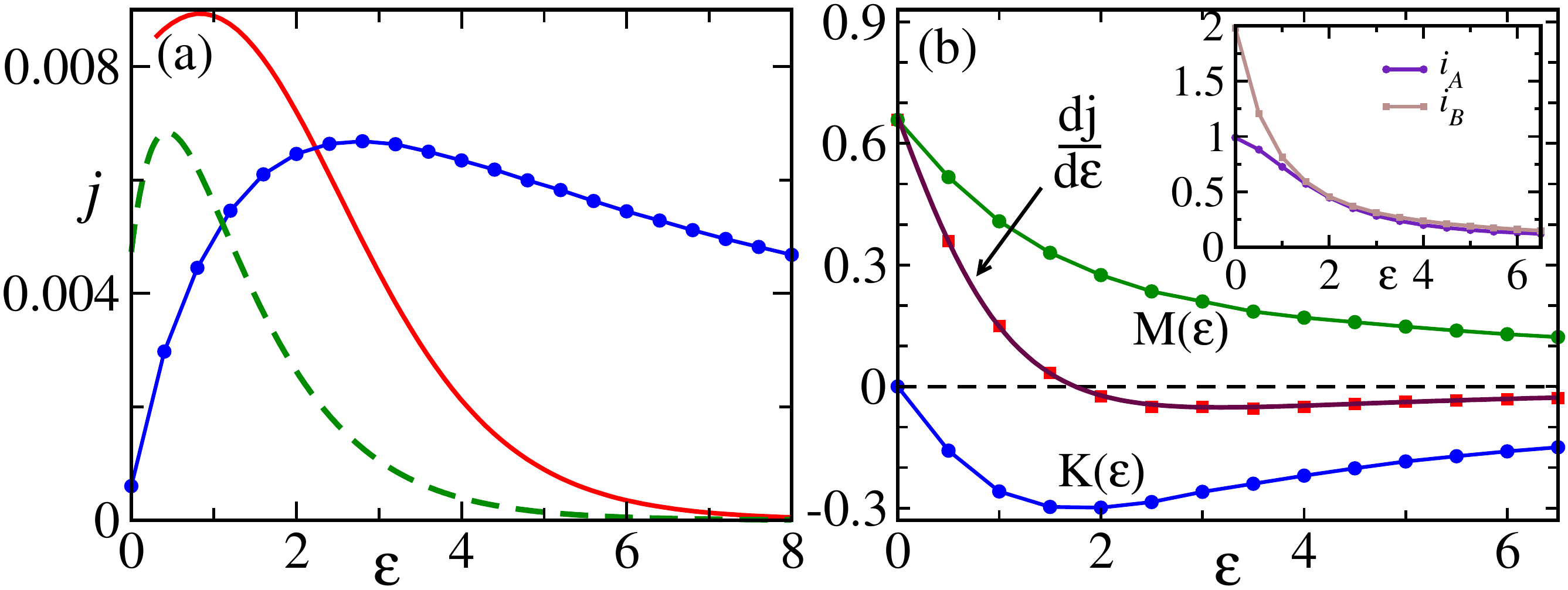}
 \caption{(a) Current $j$ versus $\ve$ for  AMP dynamics (\ref{eq:AMP}) for   density $\rho=0.15.$
 Circle: $\psi =1/(1+\ve)$ (simulations),  solid line: exact results for $\psi$ given by Eq. (\ref{eq:p0_AMP}).
 Dashed line: current  $j$ (multiplied by $0.01$ for better visibility)  for AZRP dynamics  (\ref{eq:ZRP}). (b) Entropic and frenetic components of the linear response in the TRW model as a function of the driving $\ve$ measured from Monte-Carlo simulations of a system of size $L=100.$ The inset shows the  traffic $i_A, i_B.$}\label{fig:resp}
\end{figure}

To explore the possibility of NDM  in this system  we did  Monte Carlo simulation with $\psi(\ve)=1/(1+\ve).$ Figure~\ref{fig:resp}(a) shows the particle current $j$ versus $\ve$ (symbols)
 which depicts a non-monotonic behaviour; once again we see that slowing down a non-driven mode results in a NDM.
This behaviour of current can be understood more rigorously  from  the  exact steady state weights of AMP,  which is of a factorized form  $P(\{n_i\}) \sim \prod_i f(n_i),$  when the rate functions  satisfy certain conditions~\cite{AMP}. 
In the present case, these conditions require 
\beq
\psi(\ve)=\frac{2- e^\ve +2 \delta (1-e^{-\ve})}{3 e^{\varepsilon}-4}
\label{eq:p0_AMP}
\eeq
with $\delta=\frac{1}{4} (e^{\ve}-2+\sqrt{4+12 e^{\ve}+e^{2\ve}})$,
when $f(n) = \delta^{n-1} ~\forall~ n>0$  and  $f(0)=1.$
Note that, this $\psi(\ve)$ is still a decreasing function but the model is well defined only in the regime $\ve>\ln \frac43$ where  $\psi>0.$   The grand  canonical partition function is $Z_L = [F(z)]^L$  with   $F(z) =\sum_n  f(n) z^n = 1+\frac{z}{1-\delta z},$  where  fugacity $z$ controls  the particle density through $\rho(z)= z F'(z)/F(z) = z [(1-\delta z)(1+z -\delta z)]^{-1}.$  Finally, the current is, 
\bea
j&=&\frac{1}{2 [F(z)]^2}[ (F(z)+ 2\psi-2e^{-\ve}-1 ) (F(z)-1-z) \cr
&& ~~~~~~~~~~+ 2 z(1-\psi) (F(z)-1) ].
 \eea
Fig. \ref{fig:resp}(a) shows $j$ as a function of $\ve$ for density $\rho=0.15;$ NDM  is observed   for $\ve\gtrsim 0.9$. 

The  factorized  steady state  of AMP discussed above is also a steady state  of the asymmetric   zero range process (AZRP) for a class of rate  functions \cite{AMP}, one example being  
\beq
u_r(n)=\frac{v(n)}{2}(2-v(n-1)), u_l(n)=  v(n)- u_r(n) \label{eq:ZRP}
\eeq  
where $v(n)= f(n-1)/f(n).$ We  find that current in AZRP with above dynamics also exhibits NDM for large densities; see the dashed curve in Fig. \ref{fig:resp}(a). What appears essential for the occurrence of NDM  is the asymmetric  rate functions which, we think,  can  be realized experimentally with  colloidal particles \cite{colloid} in asymmetric separate channels \cite{sepChnl, GeoAsym}.

{\it Nonequlibrium response relation:}
Away from equilibrium, the linear response of current $J$ can be expressed as a sum of two nonequilibrium correlations~\cite{Baiesi2009},
\beq
\frac{\id}{\id \ve} \la J\ra = \frac 12 \left\la S'(\omega); J \right\ra - \left \la D'(\omega); J \right\ra. \label{eq:dO}
\eeq
where $S(\omega)$ and $D(\omega)$ are the entropy (anti-symmetric under time reversal) and `frenesy' (symmetric) associated with a trajectory $\omega$ during time interval $[0,t];$ primes denote derivatives w.r.t $\ve$ and  $\la f;g\ra \equiv \la fg\ra - \la f\ra \la g\ra.$ For a single driven tracer,  $S'= J$ and  the entropic term is simply the variance of the current while the frenetic one depends on the details of the specific dynamics. A large frenetic contribution which may occur, for example, in presence of traps or obstacles, can make the overall response negative \cite{neg,neg2}. To understand how the entropic and frenetic components of mobility compete in systems where the escape rate (or the time-symmetric traffic) of the driven  particle or mode is  fixed whereas a {\it non-driven mode} is slowed down, let us consider  the example of TRW model with  $p+q=1.$  Since the driving is associated with the $A$ particle only, we have $S'= J_A,$  where $J_A$ is the time-integrated current of the $A$ particle during the time $[0,t]
.$  
The change in dynamical activity is now, 
\beq
D^\prime(\omega) = \frac {(p-q)}2 I_A - \frac{\psi^\prime}{\psi} I_B - (p^\prime+ \psi^\prime) t_{AB}  -(q^\prime + \psi^\prime) t_{BA}\n
\eeq
where $t_{AB}$($t_{BA}$) refers to the total time during which $B$ sits immediately to the right (left) of $A.$    For the stationary current $j= \lim_{t \to \infty}\la J\ra/t,$ the entropic component is then given by $M(\ve) = \lim_{t \to \infty} \la J_A; J\ra/2t$ and the frenetic component $K(\ve) = - \lim_{t \to \infty} \la D'(\omega); J \ra/t.$

Figure~\ref{fig:resp}(b) shows plots of $M(\ve)$ and $K(\ve)$ obtained from Monte Carlo simulations. Similar to the single particle case,  $M(\ve)$ remains positive for all $\ve>0$ whereas $K(\ve)$  becomes negative  resulting in NDM above a threshold field. The inset shows the average time-symmetric traffic  $i_{A,B},$ both decrease as the driving is increased; the unbiased $B$ particle, being slowed down by the field $\ve,$ in turn slows down the $A$ particle \cite{note2}.

{\it Conclusion:}
In this article, we address the question of negative differential mobility  (NDM) in interacting driven diffusive systems. It is known that NDM can occur when the driven particles are slowed down (increased time-scale of motion) by the external field. Here we propose an alternate mechanism and show that NDM can occur in multi-component system when the drive slows down some other, undriven, degree of freedom. First we illustrate this phenomenon in an exclusion process with two particle species where only one type of particles are driven by an external field. The other particles, although unaffected directly by the drive, slows down due to mutual interaction, resulting in NDM. To understand the mechanism we study a pedagogical example of two distinguishable random walkers on a periodic lattice interacting via exclusion only, one of which is driven by an external field. Other, more complex, exactly solvable examples of two-species exclusion process and asymmetric misanthrope process are also studied where the same mechanism leads to NDM for large driving. This mechanism  provides a new direction  to the  occurrence of NDM in interacting particle systems,  in contrast to the existing ones --- jamming,  kinetic constraints or trapping of driven  modes.


\begin{thebibliography}{99}

\bibitem{Kubo} R. Kubo,  Rep. Prog. Phys. {\bf 29}, 255 (1966).

\bibitem{Green} M. S. Green, J. Chem. Phys. {\bf 22},398 (1954); Phys. Rev. {\bf 119}, 829 (1960).
\bibitem{Kubo2} R. Kubo, J. Phys. Soc. Jpn. {\bf 12}, 570 (1957).


\bibitem{Baiesi2009} M. Baiesi, C. Maes and B. Wynants, 
Phys. Rev. Lett. {\bf 103}, 010602 (2009).

\bibitem{anm1} R. Eichhorn, P. Reimann, and P. H\"{a}nggi, Phys. Rev. Lett. {\bf 88}, 190601 (2002).

\bibitem{anm2} A. Ros, R. Eichhorn, J. Regtmeier, T. T. Duong, P. Reimann and D. Anselmetti,
Nature {\bf 436} 928 (2005).

\bibitem{anm3}  L. Machura, M. Kostur, P. Talkner, J. \L{}uczka, and P. H\"{a}nggi, Phys. Rev. Lett. {\bf 98}, 040601 (2007).

\bibitem{anm4} P. K. Ghosh, P. H\"{a}nggi, F. Marchesoni, and F. Nori, Phys. Rev. E {\bf 89}, 062115 (2014).

\bibitem{zia}  R. K. P.~Zia,  E.~L.~Pr{\ae}stgaard, and O.G.~Mouritsen,Am. J. Phys. {\bf 70}, 384 (2002).  

\bibitem{rev} R. Eichhorn, P. Reimann, B. Cleuren, and C. Van den Broeck, Chaos {\bf 15}, 026113 (2005).


\bibitem{Barma} M. Barma and D. Dhar, J. Phys. C {\bf 16}, 1451 (1983).

\bibitem{Jack} R. L. Jack, D. Kelsey, J. P. Garrahan, and D. Chandler, Phys. Rev. E {\bf 78}, 011506 (2008).

\bibitem{Li} B. Li, L. Wang, G. Casati, Appl. Phys. Lett. {\bf 88}, 143501 (2006).
\bibitem{Hu} B. Hu, D. He, L. Yang, and Y. Zhang, Phys. Rev. E {\bf 74}, 060101(R) (2006).
\bibitem{deb} D. Bagchi, J. Phys.: Cond. Mat. {\bf 25}, 496006  (2013). 

\bibitem{neg} P. Baerts, U. Basu, C. Maes, S. Safaverdi, Phys. Rev. E {\bf 88}, 052109 (2013).

\bibitem{Dhar} D. Dhar, J. Phys. A {\bf 17}, L257 (1984).

\bibitem{Franosch} S. Leitmann and T. Franosch, Phys. Rev. Lett. {\bf 111}, 190603 (2013).

\bibitem{neg2} U. Basu and C. Maes, J. Phys. A: Math. Theor. {\bf 47}, 255003  (2014).

\bibitem{Baiesi2015} M. Baiesi, A. L. Stella, and C. Vanderzande, Phys. Rev. E {\bf 92}, 042121 (2015).


\bibitem{Sarracino2016} A. Sarracino, F. Cecconi, A. Puglisi, and A. Vulpiani, Phys. Rev. Lett. {\bf 117} 174501 (2016).


\bibitem{Saraccino} O. B\'{e}nichou, P. Illien, G. Oshanin, A. Sarracino, and R. Voituriez, 
Phys. Rev. Lett. {\bf 113} 268002 (2014).

\bibitem{Benichou2016} O. B\'{e}nichou, P. Illien, G. Oshanin, A. Sarracino, and R. Voituriez,
Phys. Rev. E {\bf 93}, 032128(2016). 

\bibitem{Sellitto} M. Sellitto,  Phys. Rev. Lett. {\bf 101}, 048301 (2008).

\bibitem{Reichhardt} C. Reichhardt and  C. J. O. Reichhardt, J. Phys.: Condens. Matter,  {\it in press} (2017). 

\bibitem{Baiesi13} M. Baiesi, C. Maes, New J. Phys. {\bf 15}, 013004 (2013).


\bibitem{kls}
S.~Katz, J.L.~Lebowitz, and H.~Spohn, J. Stat. Phys. {\bf 34}, 497 (1984).

\bibitem{note1} We observe it for an extensive range of parameters; the data are not shown here.


\bibitem{MPAevans} R. A. Blythe, M. R. Evans, J. Phys. A Math. Theor. {\bf 40}, R333 (2007).

\bibitem{ac} M. Januszewski and J. \L{}uczka, Phys. Rev. E {\bf 83}, 051117 (2011).

\bibitem{AMP} A. K. Chatterjee and P. K. Mohanty, J. Stat. Mech.  093201 (2017).




\bibitem{colloid} R. Eichhorn, J. Regtmeier, D. Anselmettib and P. Reimann, 
 Soft Matter {\bf 6}, 1858 (2010).

 \bibitem{sepChnl} J. Wu and  B. Ai, Sc. Rep.  {\bf 6}, 24001 (2016).

 \bibitem{GeoAsym} R. S. Shaw, N. Packard, M. Schroter, and H. L. Swinney, PNAS {\bf 104}, 9580 (2007).

\bibitem{note2} Note that, in absence of any interaction between $A$ and $B,$ $i_A$ would be a constant since $p+q=1.$  
  
\end{thebibliography}
\end{document}